\documentclass[12pt]{article}

\usepackage{mathrsfs}
\usepackage{array}
\usepackage{amssymb}
\usepackage{amsmath}
\usepackage{graphicx}
\newcommand{\p}{\partial}

\newcommand{\M}{\mathscr{M}}

\newcommand{\na}{\nabla}

\def\makeatletter{\catcode`\@=11}
\makeatletter
\def\mathbox#1{\hbox{$\m@th#1$}}%
\def\math@ccstyles#1#2#3#4#5#6#7{{\leavevmode
      \setbox0\mathbox{#6#7}%
      \setbox2\mathbox{#4#5}%
      \dimen@ #3%
      \baselineskip\z@\lineskiplimit#1\lineskip\z@
      \vbox{\ialign{##\crcr
             \hfil \kern #2\box2 \hfil\crcr
             \noalign{\kern\dimen@}%
             \hfil\box0\hfil\crcr}}}}
\def\mathaccstyles{\math@ccstyles\maxdimen}
\def\maththroughstyles{\math@ccstyles{-\maxdimen}}
\def\unitmatrixDT%
 {\maththroughstyles{.45\ht0}\z@\displaystyle {\mathchar"006C}\displaystyle 1}

\begin{document}

\begin{titlepage}
\begin{center}

\rightline{UG-10-76}
\rightline{May 25, 2011}

\vskip 1.5cm

{\Large \bf \hskip -.4truecm  Newtonian Gravity  and the Bargmann Algebra}

\vskip 1cm

{\bf \hskip -1.1truecm Roel Andringa$\,{}^*$, Eric Bergshoeff\,${}^*$, Sudhakar Panda\,${}^\dagger$ and Mees de Roo\,${}^*$}

\vskip 25pt

{\em \hskip -.1truecm ${}^*$\ Centre for Theoretical Physics, University of
Groningen, Nijenborgh 4, 9747 AG Groningen, The Netherlands\vskip 5pt}

{\em \hskip -.1truecm ${}^\dagger$\ Harish-Chandra Research Institute, Allahabad-211019, India\vskip 5pt}

{email: {\tt R.Andringa@rug.nl, E.A.Bergshoeff@rug.nl, Panda@mri.ernet.in, M.de.Roo@rug.nl}}

\vskip 15pt

\end{center}

\vskip 0.5cm

\begin{center} {\bf ABSTRACT}\\[3ex]
\end{center}

We show how the Newton-Cartan formulation of Newtonian gravity can be obtained from gauging the Bargmann
algebra, i.e., the centrally extended Galilean algebra. In this gauging procedure several curvature constraints 
are imposed. These convert the spatial (time) translational symmetries of
the algebra into spatial (time) general coordinate transformations, and make the spin connection
gauge fields dependent. In addition we require two independent  Vielbein postulates for the temporal and spatial 
directions. In the final step we  impose  an additional curvature constraint to
establish the connection with (on-shell) Newton-Cartan theory.
We discuss a few extensions of our work that are relevant in the context of the AdS-CFT correspondence.



\end{titlepage}

\newpage
\setcounter{page}{1} \tableofcontents

\newpage

\setcounter{page}{1} \numberwithin{equation}{section}

\section{Introduction}

It is well known that Einstein's formulation of gravity can be obtained by performing a formal gauging
procedure of the Poincar\'{e} algebra. In this procedure one associates to each generator of the Poincar\'{e}
algebra a gauge field. Next, one imposes constraints on the curvature tensors of these gauge fields such
that the translational symmetries of the algebra get converted into general coordinate
transformations. At the same time the gauge field of the  Lorentz transformations gets expressed into
(derivatives of) the Vierbein gauge field which is the only independent gauge field. One thus obtains an off-shell formulation of Einstein gravity.
On-shell Einstein gravity is obtained by imposing the usual Einstein equations of motion.

One may consider the non-relativistic version of the  Poincar\'{e} algebra and Einstein gravity independently. It turns out that the relevant non-relativistic version of the Poincar\'{e} algebra is
a particular contraction of the Poincar\'{e} algebra trivially extended with a 1-dimensional algebra that commutes with all the generators. This contraction yields the so-called Barg\-mann algebra, which
is the centrally extended Galilean algebra.
On the other hand, taking the  non-relativistic limit of general relativity leads to the well-known
non-relativistic Newtonian gravity
in flat space. The Newton-Cartan theory is a geometric re-formulation of this Newtonian  theory, mimicking
as much as possible the geometric formulation of general relativity \cite{Cartan,MTW}. A notable difference with the
relativistic case is the occurrence of a degenerate metric.

The question we pose in this note is: can we derive the Newton-Cartan formulation of Newtonian
gravity directly from gauging the Bargmann algebra in the same way that Einstein gravity may be derived
from gauging the relativistic Poincar\'{e} algebra as described above?\,\footnote{The gauging of the
Bargmann algebra, from a somewhat different point of view, has been considered before in
\cite{Duval:1983pb,PietriLusannaPauri}.} The answer will be
yes, but there are some subtleties involved. This is  partly due  to the
fact that the standard procedure leads to  spin-connection  fields  that not only depend on the temporal and spatial Vielbeins but also on the gauge field corresponding to the
central charge generator.
These connections have to be fixed appropriately, via further curvature constraints, in order to obtain the correct non-relativistic Poisson
equation as well as the geodesic equation for a massive particle.

The outline of this note is as follows. In section 2 we first review how Einstein gravity may be obtained
by gauging the Poincar\'{e} algebra. To keep the discussion in this section as general as possible we leave the dimension
$D$ of spacetime arbitrary. Next, we briefly review in section 3 the Newton-Cartan formulation of
Newtonian gravity, since this is the theory we wish to end up with in the non-relativistic case. We
next proceed, in section 4, with gauging the Bargmann algebra. In a first step we  introduce a set of curvature constraints
that convert the spatial (time) translational symmetries of the algebra into
spatial (time) general coordinate transformations.  We next impose a Vielbein postulate for the Vielbeins in the
temporal and spatial directions.
In a final step we impose further curvature  constraints on the theory in order to recover the non-relativistic Poisson equation
and the geodesic equation for a massive particle.
Finally, our conclusions and suggestions for further work are presented in section 5.

\section{Einstein Gravity and Gauging the Poincar\'{e} Algebra}

In this section we briefly review how the basic ingredients of Einstein gravity may be obtained by applying
a formal gauging procedure to the Poincar\'{e} algebra. We leave the dimension $D$ of spacetime in this section arbitrary.

Our  starting point is the $D$-dimensional  Poincar\'{e} algebra $\mathfrak{iso}(D-1,1)$
with generators $P_a, M_{ab}\, (a=0,1,\cdots , D-1)$
\begin{align}
[P_{a},P_{b}] &=0\,, \nonumber\\[.1truecm]
[M_{bc},P_{a}] &= -2\eta_{a[b}P_{c]}\,, \nonumber\\[.1truecm]
[M_{cd},M_{ef}] &= 4 \eta_{[c [e}M_{f]d]}\,. \label{Poincarealgebra}
\end{align}
Associating a gauge field $e_{\mu}{}^a$ to the local $P$-transformations with spacetime dependent
parameters $\zeta^a(x)$, and a gauge field $\omega_{\mu}{}^{ab}$ to the local Lorentz transformations
with spacetime dependent parameters $\lambda^{ab}(x)$, we obtain the following transformation rules
\begin{align}
\delta e_{\mu}{}^a & = \p_{\mu}\zeta^a - \omega_{\mu}{}^{ab} \zeta^b + \lambda^{ab}e_{\mu}{}^b\,,
\label{etransformation}  \nonumber\\[.1truecm]
\delta \omega_{\mu}{}^{ab} & = \p_{\mu}\lambda^{ab} + 2 \lambda^{c[a}\omega_{\mu}{}^{b]c}\,.
\end{align}

In order to make contact with gravity we wish to replace the local $P$-trans\-for\-ma\-tions of all
gauge fields by general coordinate transformations and to interpret $e_{\mu}{}^a$ as the Vielbein,
with the inverse Vielbein field $e_a{}^\mu$ defined by
\begin{equation}
e_\mu{}^a\, e_b{}^\mu = \delta_b{}^a\,,\hskip 1truecm e_\mu{}^a\, e_a{}^\nu = \delta_\mu{}^\nu\,.
\end{equation}
To show how this can be achieved by imposing curvature constraints we first consider the following
general identity for a gauge algebra:
\begin{equation}
   0 = \delta_{gct}(\xi^{\lambda})B_{\mu}{}^A +
          \xi^\lambda R_{\mu\lambda}{}^A
   - \sum_{\substack{\{C\}}} \delta(\xi^{\lambda}B_{\lambda}{}^{C})B_{\mu}{}^A\,.
\label{veryimportantequation}
\end{equation}
The index $A$ labels the gauge fields and corresponding curvatures of the gauge algebra.
If we now set $A=a$ for the $P$-transformations and write the parameter $\xi^\lambda$ as
$\xi^\lambda=e_a{}^\lambda\zeta^a$
we can bring the contribution of $e_\mu{}^a$ in the sum in (\ref{veryimportantequation}) to the left-hand
side of the equation to obtain
\begin{equation}
\delta_P(\zeta^b) e_{\mu}{}^a
   =  \delta_{gct}(\xi^{\lambda})e_{\mu}{}^a + \xi^{\lambda}R_{\mu\lambda}{}^a(P)
    - \delta_M(\xi^{\lambda}\omega_{\lambda}{}^{ab})e_{\mu}{}^a\,. \label{Poincarepexchange}
\end{equation}
We see that the difference between a $P$-transformation and a general coordinate transformation is a curvature term and
a Lorentz transformation. More generally, we  deduce from the identity \eqref{veryimportantequation} that, whenever 
a gauge field transforms under a $P$-transformation, the $P$-transformations of this gauge field can be replaced by 
a general coordinate transformation plus other
symmetries of the algebra by putting the curvature of the gauge field to zero. Since the Vielbein is
the only field that transforms under the $P$-transformations, see \eqref{etransformation}, we are led to
impose the following constraint:
\begin{equation}
R_{\mu\nu}{}^a (P) = 0\,. \label{Poincarepconstraint}
\end{equation}
The same  constraint  allows us to solve for the Lorentz gauge field
$\omega_\mu{}^{ab}$ in terms of (derivatives of) the Vielbein and its inverse:
\begin{equation}
\omega_{\mu}{}^{ab}(e,\p e) = -2e^{\lambda [a} \p_{[\mu}e_{\lambda]}{}^{b]} +
e_{\mu}{}^c e^{\lambda\,a} e ^{\rho\,b } \p_{[\lambda}e_{\rho]}{}^c\,.
\end{equation}
What remains is a theory with
the Vielbein $e_{\mu}{}^a$ as the only independent field transforming under local Lorentz transformations
and general coordinate transformations and with $\omega_\mu{}^{ab}$ as the dependent spin connection field.

A $\Gamma$-connection may be introduced by imposing the Vielbein postulate:
\begin{align}
 \na_{\mu}e_{\nu}{}^a & \equiv \p_{\mu}e_{\nu}{}^a - \Gamma_{\nu\mu}^{\rho}e_{\rho}{}^a
 - \omega_{\mu}{}^{ab}e_{\nu}{}^b =0\,. \label{P}
\end{align}
The anti-symmetric part of this equation, together with the curvature constraint \eqref{Poincarepconstraint}, shows that the anti-symmetric part of the $\Gamma$-connection is zero, i.e.~there is no torsion.
From the Vielbein postulate \eqref{P} one may solve the $\Gamma$-connection in terms of the Vielbein and its
inverse as follows:
\begin{equation}
\Gamma_{\nu\mu}^{\rho} = e^{\rho}{}_a D_{\mu}e_{\nu}{}^a\,.\label{gammaomega}
\end{equation}
Here $D_\mu$ is the Lorentz-covariant derivative.  Finally, a non-degenerate metric and its inverse can
be defined as:
\begin{equation}
g_{\mu\nu} =e_\mu{}^a e_\nu{}^b\eta_{ab}\,,\hskip 1.5truecm g^{\mu\nu} = e_a{}^\mu e_b{}^\nu\eta^{ab}\,.
\end{equation}

This concludes our description of the basic ingredients of off-shell Einstein gravity and the Poincar\'{e} algebra.
These basic ingredients are an independent non-degenerate metric $g_{\mu\nu}$ and a dependent
$\Gamma$-connection $\Gamma_{\nu\mu}^\rho$ or, in the presence of flat indices, an independent Vielbein
field $e_\mu{}^a$ and a dependent spin-connection field $\omega_\mu{}^{ab}$. The theory can be put on-shell by 
imposing the Einstein equations of motion.

\section{Newton-Cartan Gravity}

From now on we restrict the discussion to $D=4$, i.e.~one time and three space directions.
We wish to review Newton-Cartan gravity as a geometric rewriting of Newtonian gravity \cite{Cartan,MTW}.  
This geometric re-formulation is motivated by the following
observation. First, consider the classical equations of motion of a massive particle,
\begin{equation}
\ddot{x}^i(t) + \frac{\p \phi(x)}{\p x^i} = 0\,, \label{classicaleom}
\end{equation}
where $x^i(t)\, (i=1,2,3)$ are the spatial coordinates, $t$ is the absolute time coordinate and
a dot indicates differentiation with respect to $t$. Furthermore,
$\phi(x^k)$ is the Newtonian potential which satisfies the Poisson equation
\begin{equation}
\p_i \p^i \phi = 4 \pi G \rho\,, \label{geometricpoisson}
\end{equation}
where $\rho$ is the mass density.
The equations of motion \eqref{classicaleom} and \eqref{geometricpoisson}
transform covariantly under the Galilei group
\begin{align}
x^0 &\rightarrow x^0 + \xi^0\,,\hskip 2truecm
x^i  \rightarrow A^i_{\ j}x^j + v^i t + d^i\,,\label{Galileitr}
\end{align}
where  $A^i_{\ j}$ is a constant group element of $\text{SO}(3)$ and   $\{v^i,d^i\}$ are three-vectors.
In addition, these equations are invariant under
\begin{align}
x^i \rightarrow x^i + a^i(t)\,,\hskip 2truecm
\phi(x) \rightarrow \phi(x) - \ddot{a}^j(t)x^j \,,
\end{align}
where $a^i(t)$ is an arbitrary time-dependent shift vector which can give rise to an acceleration.

From the Newtonian point of view the equations \eqref{classicaleom} describe a {\sl curved} trajectory in a {\sl flat} three-dimensional space. We now wish to re-interpret the same equations as a geodesic in a
{\sl curved} four-dimensional spacetime. Indeed, one may rewrite the equations
\eqref{classicaleom}  as the geodesic equations of motion
\begin{equation}
\frac{d^2 x^{\mu}}{dt^2} + \Gamma^{\mu}_{\nu\rho}\frac{dx^{\nu}}{dt}\frac{dx^{\rho}}{dt} = 0\,,
\label{geodesicequation}
\end{equation}
provided that one chooses coordinates $\{x^{\mu}\}=\{x^0,x^i\} = (t,x^i)$ and takes the
following expression for the non-zero connection fields:
\begin{align}
\Gamma^i_{00}=\delta^{ij}\p_j \phi\,,\label{connection}
\end{align}
where we have used the Euclidean three-metric. At this point $\Gamma^{\mu}_{\nu\rho}$ is a symmetric
connection independent of the metric. The coordinate choice $x^0=t$ corresponds to choosing so-called adapted
coordinates. The corresponding $D$-dimensional spacetime is called the Newton-Cartan spacetime $\M$. The
only non-zero component of the  Riemann tensor corresponding to the connection \eqref{connection} is
\begin{equation}
R^i_{\ 0j0} = \delta^{ik}\p_k \p_j \phi\,. \label{riemanntensorcomponent}
\end{equation}
If one now imposes the equations of motion $R_{00} = 4 \pi G \rho$ one obtains the Poisson equation
\eqref{geometricpoisson}. To write the Poisson equation in a covariant way we first must introduce a metric.

As it stands, the $\Gamma$-connection defined by \eqref{connection} cannot follow from a non-degenerate
four-dimensional metric. One way to see this is to consider the Riemann tensor that is defined by this
$\Gamma$-connection. The Riemann tensor, defined in terms of a metric connection based upon a non-degenerate
metric, satisfies certain symmetry properties. One may easily verify that these properties are not satisfied by the
Riemann tensor \eqref{riemanntensorcomponent}. Another way to see that a degenerate metric is unavoidable
is to consider the relativistic Minkowski metric and its inverse
\begin{equation}
\eta_{\mu\nu}/c^2 = \begin{pmatrix}
-1&0\cr
0&\unitmatrixDT_3/c^2
\end{pmatrix}             \,,\hskip 1.5truecm
\eta^{\mu\nu} = \begin{pmatrix}
-1/c^2&0\cr
0&\unitmatrixDT_3
\end{pmatrix}\,.
\end{equation}
Taking the limit $c\rightarrow\infty$ naturally leads to a degenerate covariant temporal metric $\tau_{\mu\nu}$
with three zero eigenvalues and a degenerate contra-variant spatial metric $h^{\mu\nu}$
with one zero eigenvalue.
We conclude that the Galilei group keeps  invariant {\sl two} metrics $\tau_{\mu\nu}$ and $h^{\mu\nu}$
which are degenerate, i.e.~$h^{\mu\nu}\tau_{\nu\rho}=0$. Since $\tau_{\mu\nu}$ is effectively a
$1\times 1$ matrix we will below use its Vielbein version which is defined by a covariant vector
$\tau_\mu$ defined by $\tau_{\mu\nu} =\tau_\mu\tau_\nu$.

A degenerate spatial metric $h^{\mu\nu}$ of rank $3$ and a degenerate temporal Vielbein $\tau_\mu$ of rank 1, together with a
symmetric connection $\Gamma_{\mu\nu}^{\rho}$ on $\M$, that depends on these two degenerate metrics, can
be introduced as follows \cite{Ehlers}. First of all the degeneracy implies that
\begin{equation}
h^{\mu\nu}\tau_{\nu} = 0\,. \label{zeroeigenvalue}
\end{equation}
We next impose  metric compatibility:
\begin{align}
\na_{\rho}h^{\mu\nu}  = 0\,,\hskip 2truecm
\na_{\rho}\tau_{\mu}  = 0\,. \label{metricconditions}
\end{align}
The covariant derivative $\na$ is with respect to a connection $\Gamma_{\mu\nu}^{\rho}$. The second of these
conditions indicates that
\begin{equation}
\tau_{\mu} = \p_{\mu} f(x^{\nu}) \label{taucondition}
\end{equation}
for a scalar function $f(x^{\nu})$. In Newton-Cartan theory this scalar function is chosen to be the absolute time $t$ which
foliates $\M$:
\begin{equation}
f(x^{\nu}) \equiv t\,.
\end{equation}
In general relativity metric compatibility allows one to write down the connection in terms of the metric
and its derivatives in a unique way, see eq.~\eqref{gammaomega}. In the present analysis, the connection $\Gamma_{\mu\nu}^{\rho}$ is not uniquely
determined by the metric compatibility conditions (\ref{metricconditions}). This
can be seen from the fact that  these conditions are preserved by the shift
\begin{equation}
\Gamma^{\rho}_{\mu\nu} \rightarrow \Gamma^{\rho}_{\mu\nu} + h^{\rho\lambda}K_{\lambda(\mu}\tau_{\nu)}
\label{connectionshift}
\end{equation}
for an arbitrary two-form $K_{\mu\nu}$ \cite{Dautcourt}. Using this arbitrary two-form it is possible
to write down  the most general connection which is compatible with (\ref{metricconditions}). In order to do
this, one needs to introduce  new tensors, the spatial inverse metric $h_{\mu\nu}$ and the temporal inverse Vielbein $\tau^{\mu}$ which are defined by the following properties:
\begin{alignat}{2}
h^{\mu\nu}h_{\nu\rho} & = \delta^{\mu}_{\rho} - \tau^{\mu}\tau_{\rho}, \ \ \ \ & \tau^{\mu}\tau_{\mu} = 1\,,
\nonumber\\
h^{\mu\nu}\tau_{\nu} & = 0, & h_{\mu\nu}\tau^{\nu} = 0\,.
\label{tensorproperties}
\end{alignat}
Note that from these conditions it follows that
\begin{equation}
\na_{\rho}h_{\mu\nu} = -2\tau_{(\mu}h_{\nu)\sigma}\na_{\rho}\tau^{\sigma}
\end{equation}
which is not zero in general.
The most general connection compatible with (\ref{metricconditions}) is then \cite{Dautcourt}
\begin{align}
\Gamma^{\sigma}_{\mu\nu} & = \tau^{\sigma}\p_{(\mu}\tau_{\nu)} +
\frac{1}{2}h^{\sigma\rho} \Bigl(\p_{\nu}h_{\rho\mu} + \p_{\mu}h_{\rho\nu} - \p_{\rho}h_{\mu\nu}\Bigr)
+ h^{\sigma\lambda}K_{\lambda(\mu}\tau_{\nu)}\,.
\label{covariantconnection}
\end{align}

We note that the original independent connection \eqref{connection} is quite different from the metric connection
defined in \eqref{covariantconnection}. Nevertheless, given extra conditions
discussed below, the Newton-Cartan theory with the metric connection \eqref{covariantconnection} reproduces
Newtonian gravity. To see how this goes,  it is convenient to use adapted coordinates
$x^0=t$. The conditions (\ref{taucondition}) and (\ref{tensorproperties}) then imply
\begin{align}
\tau_{\mu} & = \delta_{\mu}^0, \ \ \ \ \tau^{\mu} = (1,\tau^i)\,, \nonumber\\
h^{\mu0} & = 0, \ \ \ \ h_{\mu 0} = - h_{\mu i}\tau^i\,. \label{adaptedcoordinates}
\end{align}
These conditions are preserved by the coordinate transformations
\begin{align}
x^0 & \rightarrow x^0 + \xi^0\,, \nonumber\\
x^i & \rightarrow x^i + \xi^i (x^{\mu})\,, \label{preserved1}
\end{align}
where $\xi^0$ is a constant. The finite spatial transformation generated by $\xi^i(x^{\mu})$ is invertible.
In adapted coordinates the connection coefficients (\ref{covariantconnection}) are given by \cite{Dautcourt}
\begin{align}
\Gamma^{i}_{00} & = h^{ij}(\p_0 h_{j0} - \tfrac{1}{2}\p_j h_{00} + K_{j0}) \equiv h^{ij}\Phi_j\,, \nonumber\\
\Gamma^{i}_{0j} & = h^{ik}(\tfrac{1}{2}\p_0 h_{jk} + \p_{[j} h_{k]0} - \tfrac{1}{2}K_{jk})
\equiv h^{ik} (\tfrac{1}{2}\p_0 h_{jk} + \omega_{jk})\,, \nonumber\\
\Gamma^{i}_{jk} & = \{^{\ i}_{jk}\}\,, \hskip 2truecm
\Gamma^{0}_{\mu\nu}  = 0\,, \label{adaptedconnection}
\end{align}
where $\{^{\ i}_{jk}\}$ are the usual Christoffel symbols with respect to the metric $h_{ij}$ with
inverse $h^{ij}$.

We now replace the original equations of motion $R_{00}=4 \pi G \rho$ by the covariant Ansatz
\begin{equation}
R_{\mu\nu} = 4 \pi G \rho\, \tau_{\mu}\tau_{\nu} \label{covarianteom}
\end{equation}
and verify that this leads to Newtonian gravity. In adapted coordinates these equations imply that
\begin{equation}
R_{ij}=R_{i0}=0\,.
\end{equation}
The condition $R_{ij}=0$ implies that the spatial
hypersurfaces are flat, i.e.~one can choose a coordinate frame with $\Gamma^i_{jk}=0$ such that the
spatial metric is given by
\begin{align}
h_{ij}  = \delta_{ij}, \ \ \ \ h^{ij}  = \delta^{ij}\,.
\end{align}
This  implies
\begin{align}
\Gamma^{i}_{0j} & = h^{ik}\omega_{jk} \ \ \leftrightarrow \ \ \omega_{ij} = h_{k[j} \Gamma^k_{i]0}\,, \nonumber\\
\Gamma^{i}_{00} & = h^{ij}\Phi_j\  \ \ \leftrightarrow\ \ \ \Phi_i = h_{ij}\Gamma^j_{00}\,. \label{omegaphigamma}
\end{align}
The choice of a flat metric further reduces the allowed coordinate transformations (\ref{preserved1}) to
\begin{align}
x^0 & \rightarrow x^0 + \xi^0\,,\hskip 2truecm
x^i  \rightarrow A^{i}_{\ j}(t) x^j + a^i(t),
\label{preserved2}
\end{align}
where $A^i_{\ j}(t)$ is an element of $\text{SO}(3)$.

To derive the Poisson equation from the Ansatz \eqref{covarianteom} two additional conditions must be invoked.
The first is the Trautman condition \cite{Trautman}:
\begin{equation}
h^{\sigma [\lambda}R^{\mu]}_{\ (\nu\rho)\sigma}(\Gamma) = 0\,.
\label{Trautman1}
\end{equation}
In adapted  coordinates it implies
\begin{align}
\p_0\omega_{mi} - \p_{[m}\Phi_{i]} = 0\,,\hskip 2truecm  \p_{[k}\omega_{mi]} = 0\,.
\label{Trautman2}
\end{align}
Although $\Phi_i$ and $\omega_{ij}$ are not tensors, both equations of (\ref{Trautman2}) are separately
covariant under (\ref{preserved2}) which can be checked explicitly. Using the definitions
\eqref{omegaphigamma} of $\Phi_i$ and $\omega_{ij}$ one may verify that the
conditions (\ref{Trautman2}) are equivalent to the manifestly tensorial equation
\begin{equation}
\p_{[\rho}K_{\mu\nu]} = 0 \ \ \rightarrow \ \ K_{\mu\nu} = 2\p_{[\mu}m_{\nu]}\,,\label{Trautman3}
\end{equation}
where $m_{\mu}$ is a vector field determined up to the derivative of some scalar field.

The second condition we need is that $\omega_{ij}$, see  (\ref{adaptedconnection}), depends only on time,
not on space coordinates \cite{Ehlers,Dautcourt}. In \cite{Ehlers} three possible conditions on the Riemann tensor
are discussed that lead to the desired restriction on $\omega_{ij}$:
\begin{equation}
h^{\rho\lambda} R^{\mu}_{\ \nu\rho\sigma}(\Gamma) R^{\nu}_{\ \mu\lambda\alpha}(\Gamma)  = 0 \ \ \text{or} \ \
\tau_{[\lambda}R^{\mu}_{\ \nu ] \rho\sigma}(\Gamma)  = 0 \ \ \text{or} \ \
h^{\sigma [\lambda}R^{\mu]}_{\ \nu\rho\sigma}(\Gamma)  = 0. \label{ehlersconditions}
\end{equation}
These are the so-called Ehlers conditions.
Each condition separately leads to the condition $\p_k\omega_{ij} = 0$ in adapted coordinates and thus $\omega_{ij}=\omega_{ij}(t)$.
One can next set $\omega'_{ij}\equiv 0$, or equivalently $\Gamma^{'i}_{0j} \equiv 0$, see \eqref{omegaphigamma},  by a time-dependent rotation $x^{'i}= A^i_{\ j}(t)x^j$  \cite{Dautcourt}.
The conditions (\ref{Trautman2}) imply that in the new coordinate
system $\partial_{[i}^\prime \Phi_{j]}^\prime=0$ and hence that $\Phi_i' = \partial_i' \Phi$ for some
scalar field $\Phi$. This implies that
\begin{equation}
\Gamma^{'i}_{00} = \delta^{ij}\partial^{'}_{j} \Phi
\end{equation}
in this coordinate system. The equations (\ref{covarianteom}) thus
lead to the Poisson equation:
\begin{align}
R_{00} & = \p_i \Gamma^i_{00} =\delta^{ij} \p_i \p_j \phi = 4 \pi G \rho\,. \label{Poisson2}
\end{align}

Finally, we should also recover the geodesic equation (\ref{geodesicequation}). Using adapted coordinates
and performing the above time-dependent rotation indeed gives the desired equations:
\begin{equation}
\ddot{x}^{'0}(t) = 0, \hskip 2truecm \ddot{x}^{'i}(t) + \partial^{'i} \Phi = 0\,.
\end{equation}
This completes the proof that Newton-Cartan gravity, formulated in terms of two degenerate metrics
(see eq.~\eqref{zeroeigenvalue}), and supplied with the Trautman condition \eqref{Trautman1}
and the Ehlers conditions \eqref{ehlersconditions}, precisely leads to the equations of Newtonian gravity. In the next
section we will show how the same Newton-Cartan theory, including the Trautman and Ehlers conditions,  follows
from gauging the so-called Bargmann algebra.

\section{Gauging the Bargmann algebra}

\subsection{The Bargmann algebra}

The Bargmann algebra is the Galilean algebra augmented with a central generator\,\footnote{In
$D=3$ dimensions three such central generators can be introduced \cite{Brihaye:1995nv,Bonanos:2008kr}.}
$M$ and can be obtained as follows. We first extend the Poincar\'e algebra $\mathfrak{iso}(D-1,1)$ to
the direct sum of the Poincar\'{e} algebra and a commutative subalgebra $\mathfrak{g}_M$ spanned by $M$:
\begin{equation}
\mathfrak{iso}(D-1,1) \ \ \rightarrow\ \  \mathfrak{iso}(D-1,1) \oplus \mathfrak{g}_M\,. \label{directsumgalilei}
\end{equation}
We next perform the following contraction of this algebra:
\begin{align}
P_0 & \rightarrow \frac{1}{\omega^2}M + H\,, \hskip 1truecm 
P_i  \rightarrow \frac{1}{\omega}P_i\,, \hskip 1truecm 
J_{i0}  \rightarrow \frac{1}{\omega}G_i\,,\hskip 1truecm \omega \rightarrow 0\,.\label{contraction}
\end{align}
The contraction of $P_0$ is motivated by considering the non-relativistic approximation of $P_0$ for  a massive free particle
\begin{align}
P_0 & = + \sqrt{c^2 P_i P^i + M^2c^4}
\ \approx\ Mc^2 + \frac{P_i P^i}{2M}\,,
\end{align}
where $c = \omega^{-1}$ is the speed of light.
The contracted algebra is the so-called Bargmann algebra $\mathfrak{b}(D-1,1)$ which has the following
non-zero commutation relations:
\begin{align}
[J_{ij},J_{kl}] & = 4 \delta_{[i[k}J_{l]j]}\,, \hskip 1truecm
[J_{ij},P_k]  = -2\delta_{k[i}P_{j]}\,, \nonumber\\
[J_{ij}, G_k] & = -2 \delta_{k[i}G_{j]}\,,\hskip 1truecm
[G_i, H]  = - P_i\,, \nonumber\\
[G_i,P_j] & = -\delta_{ij}M \,,\label{Bargmannalgebra}
\end{align}
For $M=0$ this is the Galilean algebra.

The $M$ generator is needed to obtain massive representations of the Galilean algebra. This can
be understood by considering the action for a non-relativistic free particle with mass $M$:\,\footnote{We thank J.~Gomis for showing this argument to us.}
\begin{equation}
S = \frac{1}{2} \int_{t_1}^{t_2} M\dot{x}^i \dot{x}^i dt\,.
\end{equation}
This action is invariant under the Galilei transformations \eqref{Galileitr}, but the Lagrangian
$L$ is not; it transforms as a total derivative under an infinitesimal Galilei boost $\delta x^i = v^i t$:
\begin{equation}
\delta L = \frac{d}{dt}\Bigl(M x^i v^i\Bigr)\,.
\end{equation}
Due to this the naive Noether charge  $Q_{\text{naive}}=p^i\delta x^i = M\dot{x}^i v^i t$ gets
modified by an additional boundary term such that the correct Noether charge corresponding to boosts becomes:
\begin{equation}
Q_G = M\dot{x}^i v^i t - M x^i v^i\,.
\end{equation}
Using this expression one may verify that the Poisson bracket of the Noether charge $Q_G$ corresponding to  infinitesimal
boosts $\delta x^i = v^i t$ with the Noether charge $Q_P$ corresponding to infinitesimal translations
$\delta x^i = a^i$ indeed gives the
central generator $M$:
 \begin{align}
\{Q_G, Q_P \}_{PB} =
   - Mv^k a^k\,,
\end{align}
in line with the $[G_i,P_j]$ commutator given in \eqref{Bargmannalgebra}.

\subsection{Gauging the Bargmann algebra}

We now gauge the Bargmann algebra \eqref{Bargmannalgebra} following  the same procedure we applied
to  the Poincar\'{e} algebra \eqref{Poincarealgebra} in Section 2.

Compared to the Poincar\'{e} case the gauge fields and parameters corresponding to the Bargmann
algebra split up  into a spatial and temporal part:
\begin{align}
e_{\mu}{}^a & \rightarrow \{e_{\mu}{}^0, \  e_{\mu}{}^i \}\,,\hskip 2truecm
\omega_{\mu}{}^{ab}  \rightarrow \{\omega_{\mu}{}^{ij}, \  \omega_{\mu}{}^{i0}\} \nonumber\\
\zeta^{a} & \rightarrow \{\zeta^0, \ \zeta^i\}\,,\hskip 2truecm
\lambda^{ab}  \rightarrow \{\lambda^{i0}, \ \lambda^{ij}\}\,. \label{splitting}
\end{align}
The gauge field corresponding to the generator $M$ will be called $m_{\mu}$ and its gauge parameter will
be called $\sigma$. We label $e_{\mu}{}^0 = \tau_{\mu}$ and $\zeta^0 = \tau$. The variations of
the gauge fields corresponding to the different generators are given by:
\begin{align}
H: \ \ \delta \tau_{\mu} & = \p_{\mu}\tau\,, \nonumber\\
P: \ \ \delta e_{\mu}{}^i & = D_{\mu}\zeta^i + \lambda^{ij}e_{\mu}{}^j + \lambda^{i0}\tau_{\mu} -
\tau\omega_{\mu}{}^{i0}\,, \nonumber\\
G: \ \ \delta \omega_{\mu}{}^{i0} & = D_{\mu}\lambda^{i0} + \lambda^{ij}\omega_{\mu}{}^{j0}\,, \nonumber\\
J: \ \ \delta\omega_{\mu}{}^{ij} & = D_{\mu}\lambda^{ij}\,, \nonumber\\
M: \ \ \delta m_{\mu} & = \p_{\mu}\sigma - \zeta^i\omega_{\mu}{}^{i0} + \lambda^{i0}e_{\mu}{}^i\,.
\label{Galileivariations}
\end{align}
The derivative $D_{\mu}$ is covariant with respect to the $J$-transformations and as such only contains
the $\omega_{\mu}{}^{ij}$ gauge field. The curvatures of the gauge fields read
\begin{align}
R_{\mu\nu}(H) & = 2\p_{[\mu}\tau_{\nu]}\,, \\
R_{\mu\nu}{}^i (P)  & = 2(D_{[\mu}e_{\nu]}{}^i - \omega_{[\mu}{}^{i0}\tau_{\nu]})\,, \label{Pcurvature}\\
R_{\mu\nu}{}^{ij}(J)  & = 2(\p_{[\mu}\omega_{\nu]}{}^{ij} - \omega_{[\mu}{}^{ki}\omega_{\nu]}{}^{jk})\,, \\
R_{\mu\nu}{}^{i0} (G) & = 2 D_{[\mu}\omega_{\nu]}{}^{i0}\,, \\
R_{\mu\nu}(M)  & = 2(\p_{[\mu}m_{\nu]} + e_{[\mu}{}^{j}\omega_{\nu]}{}^{j0})\,. \label{Mcurvature}
\end{align}

Using the general formula (\ref{veryimportantequation}) we convert the $P$ and $H$ transformations
into general coordinate transformations in space and time. We write the parameter of the general
coordinate transformations $\xi^\lambda$ in \eqref{veryimportantequation} as
\begin{equation}
\xi^{\lambda} = e^{\lambda}{}_i \zeta^i + \tau^{\lambda}\tau\,.
\end{equation}
Here we have used the inverse spatial Vielbein  $e^{\lambda}{}_i$ and the inverse temporal Vielbein $\tau^\lambda$ defined by
\begin{align}
e_{\mu}{}^i e^{\mu}{}_j & = \delta_j^i, \ \ \ \ \ \tau^{\mu}\tau_{\mu} = 1\,, \label{constraintone} \\
\tau^{\mu} e_{\mu}{}^i & = 0, \ \ \ \ \ \tau_{\mu}e^{\mu}{}_i = \label{constrainttwo} 0\,, \\
e_{\mu}{}^i e^{\nu}{}_i & = \delta^{\nu}_{\mu} - \tau_{\mu}\tau^{\nu}\,. \label{constraintthree}
\end{align}
These conditions are the Vielbein version of the  conditions  (\ref{tensorproperties}).\\

We observe that only the gauge fields $e_{\mu}{}^i\,,\tau_\mu$ and $m_{\mu}$ transform under the $P$
and $H$ transformations. These are the fields that should remain independent, while the spin connections
should become dependent fields. This can be achieved with the following
constraints:
\begin{equation}
R_{\mu\nu}{}^i(P)  = R_{\mu\nu}(H) = R_{\mu\nu}(M) = 0 \,. \label{GalileiPHconstraints}
\end{equation}
The Bianchi identities then lead to additional relations between curvatures:
\begin{align}
R_{[\lambda\mu}{}^{ij}(J)e_{\nu]}{}^j &=  -R_{[\lambda\mu}{}^{i0}(G)\tau_{\nu]}\,,\hskip 2truecm
e_{[\lambda}{}^{i}R_{\mu\nu]}{}^{i0}(G) = 0\,.\label{furtherconstraints}
\end{align}

The constraint $R_{\mu\nu}(H)=0$ gives  the condition $\partial_{[\mu}\tau_{\nu]}=0$ and hence we
may take $\tau_\mu$ as in (\ref{taucondition}). The other two constraints, $R_{\mu\nu}{}^i(P)  =  R_{\mu\nu}(M)=0$,
enable us to solve for  the spin connection gauge fields $\omega_\mu{}^{ij}, \omega_\mu{}^{i0}$
in terms of the other gauge fields, so that indeed only $e_{\mu}{}^i\,,\tau_\mu$ and $m_{\mu}$ remain as 
independent fields.

To solve for $\omega_{\mu}{}^{ij}$, we write
\begin{equation}
R_{\mu\nu}{}^i(P)e_{\rho}{}^i + R_{\rho\mu}{}^i(P)e_{\nu}{}^i - R_{\nu\rho}{}^i(P)e_{\mu}{}^i = 0\,.
\end{equation}
From this it follows that
\begin{align}
\omega_{\mu}{}^{kl}
   & =  \partial_{[\mu}e_{\nu]}{}^k e^{\nu\,l} - \partial_{[\mu}e_{\nu]}{}^l e^{\nu\,k}
         + e_{\mu}{}^i \partial_{[\nu}e_{\rho]}{}^i  e^{\nu\,k}e^{\rho\,l}
       - \tau_{\mu}e^{\rho\,[k}\omega_{\rho}{}^{l]0} \,.
\label{solutionofomega}
\end{align}
Next we solve for $\omega_{\mu}{}^{i0}$. We substitute (\ref{solutionofomega}) into 
$R_{\mu\nu}{}^i(P)=0$ and contract this with $e^{\mu}{}_j$ and $\tau^{\nu}$.  This gives the condition
\begin{equation}
    e^{\mu\, (i}\omega_{\mu}{}^{j)0} = 2\, e^{\mu\, (i} \partial_{[\mu}e_{\nu]}{}^{j)}\tau^{\nu}\,.
\label{uno}
\end{equation}
Furthermore, $R_{\mu\nu}(M)=0$ can be contracted with $e^{\mu}{}_i$ and $\tau^{\mu}$
to give the following conditions:
\begin{align}
  e^{\mu\,[i}\omega_{\mu}{}^{j]0} & = e^{\mu\, i} e^{\nu \,j} \p_{[\mu}m_{\nu]}\,,\hskip 2truecm
 \tau^{\mu}\omega_{\mu}{}^{i0}  =  2 \tau^{\mu} e^{\nu\, i} \p_{[\mu}m_{\nu]}\,.\label{tres}
\end{align}
Using the constraints  (\ref{uno}) and (\ref{tres}) one arrives at the following solution
for  $\omega_{\mu}{}^{i0}$:
\begin{align}
\omega_{\mu}{}^{i0} & = e^{\nu i}\p_{[\mu}m_{\nu]} + e^{\nu i}\tau^{\rho} e_{\mu}{}^{j}\p_{[\nu}e_{\rho]}{}^{j} 
+ \tau_{\mu}\tau^{\nu}e^{\rho i}\p_{[\nu}m_{\rho]} + \tau^{\nu}\p_{[\mu}e_{\nu]}{}^{i}
  .\label{omegai0}
\end{align}
At this point we are left with  the independent fields  $e_{\mu}{}^i$, $\tau_{\mu}$ and $m_{\mu}$.
Furthermore, the theory is still off-shell; no equations of motion have been imposed.

\subsection{Newton-Cartan Gravity}

To make contact with the  formulation of Newton-Cartan gravity presented in Section 3 we need to
introduce a $\Gamma$-connection. In the gauge algebra approach this is most naturally done by
imposing a Vielbein postulate for the spatial Vielbein
\begin{equation}
\p_{\mu}e_{\nu}{}^i - \omega_{\mu}{}^{ij}e_{\nu}{}^j -\omega_{\mu}{}^{i0}\tau_{\nu} -
\Gamma_{\nu\mu}^{\rho}e_{\rho}{}^i =0 \label{Galileanvielbeinpostulate}
\end{equation}
and a Vielbein postulate for the temporal Vielbein
\begin{equation}
  \partial_\mu \tau_\nu - \Gamma^\lambda_{\nu\mu}\tau_\lambda = 0\,,\label{second}
\end{equation}
which is the second condition of (\ref{metricconditions}). These Vielbein postulates imply
\begin{equation}
\Gamma^{\rho}_{\nu\mu} = \tau^{\rho}\p_{(\mu}\tau_{\nu)} + e^{\rho}{}_i \Bigl( \p_{(\mu}e_{\nu)}{}^i - \omega_{(\mu}{}^{ij}e_{\nu)}{}^j -\omega_{(\mu}{}^{i0}\tau_{\nu)}	\Bigr)\,. \label{gammagaugefields}
\end{equation}
This connection is symmetric due to the curvature constraints $R_{\mu\nu}{}^i(P)= R_{\mu\nu}(H)= 0$, and satisfies 
\eqref{metricconditions}.
An important difference between the metric compatibility conditions given in (\ref{metricconditions}) and 
in (\ref{Galileanvielbeinpostulate}, \ref{second})
is that the latter define the connection $\Gamma$ uniquely.
From (\ref{covariantconnection}) and (\ref{gammagaugefields}) we find that
\begin{equation}
K_{\mu\nu} = 2 \omega_{[\mu}{}^{i0}e_{\nu]}{}^i\,,\label{K}
\end{equation}
with $\omega_\mu{}^{i0}$ given by \eqref{omegai0}. This implies via the $R(M)=0$ constraint that
\begin{equation}
K_{\mu\nu} = 2\partial_{[\mu}m_{\nu]}
\end{equation}
which solves the  condition \eqref{Trautman3}. The Riemann tensor corresponding to (\ref{gammagaugefields})
can now be expressed in terms of the curvature tensors of the gauge algebra:
\begin{align}
R^{\mu}_{\ \nu\rho\sigma}(\Gamma) & = \p_{\rho}\Gamma^{\mu}_{\nu\sigma} - \p_{\sigma}\Gamma^{\mu}_{\nu\rho} + 
\Gamma^{\lambda}_{\nu\sigma}\Gamma^{\mu}_{\lambda\rho} - \Gamma^{\lambda}_{\nu\rho}\Gamma^{\mu}_{\lambda\sigma} \nonumber\\
& = -e^{\mu}{}_i \Bigl( R_{\rho\sigma}{}^{i0}(G)\tau_{\nu}+ R_{\rho\sigma}{}^{ij}(J)e_{\nu j} \Bigr)     
\,. \label{riemanngaugefields}
\end{align}
Here we have used (\ref{GalileiPHconstraints}). The Trautman condition \eqref{Trautman1}, applied to 
\eqref{riemanngaugefields}, is equivalent to  the first constraint of \eqref{furtherconstraints}.

We know from the analysis in section 3 that, in order to make contact with the Newton-Cartan formulation, we must impose 
the Ehlers conditions \eqref{ehlersconditions}. One can show that each of the three Ehlers conditions (\ref{ehlersconditions}) is
equivalent to the single curvature constraint
\begin{equation}
R_{\mu\nu}{}^{ij}(J) = 0\,. \label{RJconstraint}
\end{equation}
Substituting this result into \eqref{furtherconstraints} leads to the following
constraints on $R_{\mu\nu}{}^{i0}(G)$:
\begin{align}
R_{[\lambda\mu}{}^{i0}(G)\tau_{\nu]} & = 0\,,\hskip 2truecm
e_{[\lambda}{}^{i}R_{\mu\nu]}{}^{i0}(G) = 0\,. \label{BianchisRG1}
\end{align}
The contraction of (\ref{BianchisRG1}) with $e^{\mu}{}_i$ and $\tau^{\mu}$ gives 
\begin{align}
e^{\mu}{}_i e^{\nu}{}_j R_{\mu\nu}{}^{k0}(G) & = 0\,,\hskip 2truecm
\tau^{\mu}e^{\nu\,[i}R_{\mu\nu}{}^{j]0}(G) = 0\,. \label{BianchisRG2}
\end{align}
This implies that the only non-zero component of $R_{\mu\nu}{}^{i0}(G)$ is
\begin{align}
\tau^{\mu}e^{\nu\ (i}R_{\mu\nu}{}^{j)0}(G) & = \delta^{k(j}R^{i)}_{\ \ 0k0}(\Gamma)
\label{finalresult}
\end{align}
which is precisely the only non-zero component (\ref{riemanntensorcomponent}) of the Riemann tensor
that occurs in the Newton-Cartan formulation.

At this point we have made contact with the Newton-Cartan gravity theory presented in Section 3. We
have the same $\Gamma$-connection and (degenerate) metrics. It can be shown that these lead to the
desired Poisson equation and geodesic equation of a massive free particle following
the same steps as in Section  3. This concludes our discussion of the gauging procedure.

\section{Conclusions}

In this work we have shown how, just like Einstein gravity, the Newton-Cartan formulation of Newtonian 
gravity can be obtained by a gauging procedure. The Lie algebra underlying this procedure is the Bargmann 
algebra  given in \eqref{Bargmannalgebra}. To obtain the correct Newton-Cartan formulation we need 
to impose constraints on the curvatures.  In a first step we impose the curvature constraints 
\eqref{GalileiPHconstraints}. They enable us to convert the spatial (time) translational symmetries 
of the Bargmann algebra into spatial (time) general coordinate transformations.
At the same time they enable us to solve for
the spin-connection gauge fields $\omega_{\mu}{}^{i0}$ and $\omega_{\mu}{}^{ij}$ in terms of the
remaining gauge fields $e_\mu{}^i\,, \tau_\mu$ and $m_\mu$, see eqs.~\eqref{solutionofomega} and \eqref{omegai0}.
For this to work it is essential that we work with a
non-zero central element $M$ in the algebra. Sofar, we work off-shell without comparing
equations of motion.

In a second step we impose  the Vielbein postulates \eqref{Galileanvielbeinpostulate} and
\eqref{second}. These enable us to solve for the $\Gamma$ connection thereby solving
the Trautman condition \eqref{Trautman1} automatically. In order to obtain the correct
Poisson equation and geodesic equation of a massive free particle we impose in a third step
the additional curvature constraints \eqref{RJconstraint} which are equivalent to each
of the three Ehlers conditions \eqref{ehlersconditions}.
The Poisson equation and the geodesic equation for a massive particle are obtained from the
relation (\ref{finalresult}) between the  curvature of the dependent field $\omega_{\mu}{}^{i0}$ 
and the Newton-Cartan Riemann tensor in the form (\ref{riemanntensorcomponent}).
The independent gauge fields $e_\mu{}^i$ and 
$\tau_\mu$ describe the degenerate metrics of Newton-Cartan gravity.

The present work can be extended in several directions. First of all, it would be
interesting to see whether a supersymmetric version of the Bargmann algebra leads to
the  Newtonian version of a Poincar\'{e} supergravity model. Secondly, one could try to apply the
gauging procedure developed in this paper to other algebras which have appeared
in recent non-relativistic applications of the AdS-CFT correspondence. Examples of such algebras are
the Galilean Conformal algebra, the Schrodinger algebra and the Lifshitz algebra. The gauging
of the first algebra is expected to lead to a Newtonian version of conformal gravity.
Irrespective of its of its role in the AdS/CFT correspondence it would be interesting to 
see whether this could lead to a non-relativistic version
of the conformal tensor calculus.

One of the original motivations of this work was the possible role of Newton-Cartan
gravity in non-relativistic applications of the AdS-CFT correspondence. In most applications
the relativistic symmetries of the AdS bulk theory are broken by the vacuum solution one
considers\footnote{For other aspects of Newton-Cartan gravity, see, e.g., 
\cite{Lin:2008pi,Duval:2009vt}}. This is the case if one
considers the Schrodinger or Lifshitz algebras. The situation changes if one considers  the Galilean 
Conformal Algebra instead.
It has been argued that in that case the bulk gravity theory is given by an extension
of the Newton-Cartan theory where the spacetime metric is degenerate with {\sl two}
zero eigenvalues corresponding to the time and the radial directions \cite{Bagchi:2009my}.
This leads to
a foliation where the time direction is replaced by a two-dimensional $\text{AdS}_2$ space.
This requires a contraction of the Poincar\'{e} algebra 
in which the Bargmann algebra is replaced by a centrally extended string Galilean algebra or,
if one includes the cosmological constant, by a string Newton-Hooke algebra 
\cite{Brugues:2004an,Gomis:2005pg}\footnote{For other applications of the Newton-Hooke algebra see, 
e.g., \cite{Tian:2004ya,Papageorgiou:2009zc}.}.
We expect that the systematic gauging procedure developed in this work will be essential to work
out the non-relativistic theories corresponding to these new cases.

\section*{Acknowledgements}

We wish to thank for useful discussions  G.~Dautcourt and, especially, J.~Gomis who clarified several 
issues in non-relativistic gravity to us. The work of R. Andringa  is supported by
an Ubbo Emmius Fellowship. S. Panda thanks the Centre for Theoretical Physics, Groningen for its
hospitality.


\end{document}